\documentstyle[11pt,newpasp,twoside,epsf]{article}
\begin{document}
\title{G28.17+0.05, A Giant Atomic/Molecular Cloud}
\author{Anthony Minter \& Felix J. Lockman}
\affil{NRAO, P.O. Box 2, Green Bank, WV 24944}

\begin{abstract}
HI 21 cm observations with the NRAO 140 Foot telescope have 
revealed a giant HI cloud (G28.17+0.05) in the Galactic 
plane that has unusual properties. The cloud is 150 pc in 
diameter, is at a distance of 5 kpc, and contains as much as 
10$^5 {\rm M_{\sun}}$ of atomic hydrogen. The cloud consists of a 
cold core, $T \sim 10~K$, and a hotter outer envelope, $T 
\ge 200~K$. There is no observable difference in the HI line 
widths, $\sim 7$ km~s$^{-1}$, between the core and the 
envelope. Anomalously-excited 1720 MHz OH emission, with a 
similar line width, is associated with the core of the 
cloud. The cloud core also exhibits $^{12}$CO and $^{13}$CO self-absorption
 which indicates that most of the cloud mass is in molecules. The 
total mass of the cloud is $> 2 \times 10^5 {\rm M_{\sun}}$. The 
cloud has only a few sites of current star formation. If 
similar clouds are associated with other observed sites of 
anomalously-excited 1720 MHz OH emission, there may be as
many as 100 more of these objects in the inner galaxy.
\end{abstract}

\section{G28.17+0.05 In HI}

G28.17+0.05 was discovered in 21cm HI observations made with the NRAO 140 Foot
Telescope (Minter et al. 2001).  At the time of these observations these data 
comprised the most
detailed look at this part of the Galactic Plane in HI.

G28.17+0.05 is shown in Figure 1.  It appears as a shell of bright emission
surrounding an area of reduced emission centered on 
${\rm l,b} = 28\fdg17+0\fdg05$.  G28.17+0.05 is estimated to have a 
distance of $\sim 5~{\rm kpc}$ at which its diameter would be $150~{\rm pc}$.
The reduced emission in the central area of 
G28.17+0.05 (hereafter referred to as the core) 
is the result of HI self-absorption (see Figure 2).  Models of
the 21cm HI emission from G28.17+0.05 indicate the temperature in the
core is ${\rm T \leq 40~K}$ while the shell has ${\rm T \geq 200~K}$.  The
core contains $1/3$ of the total HI mass.  The column density
of atomic hydrogen through the center is 
${\rm N_{HI} \geq 5-12 \times 10^{20}~cm^{-2}}$, depending on the exact
value of the temperature in the core.  The total atomic hydrogen
mass within G28.17+0.0 is $\sim 10^5 {\rm M_{\sun}}$.  The line widths (FWHM)
of HI in the core and shell are both $\sim 7~{\rm km~s^{-1}}$.
This results in G28.17+0.05 being gravitationally bound from just its atomic
mass alone!

\begin{figure}
\plotfiddle{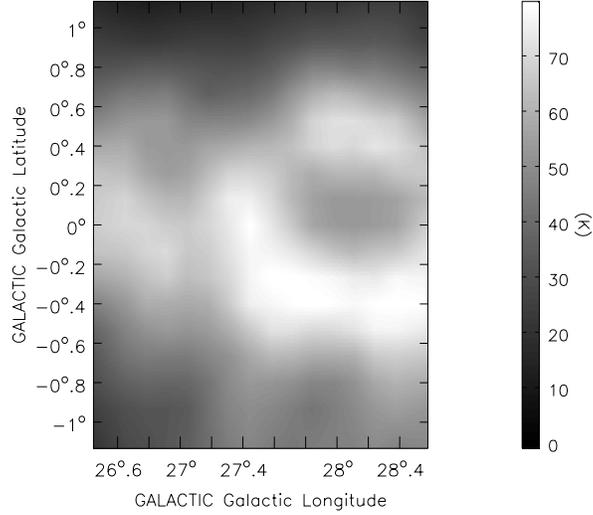}{3in}{0}{50}{50}{-140}{-90}
\caption{G28.17+0.05 21cm HI brightness temperature at 
$v_{LSR} = 77~{\rm km~s^{-1}}$.}
\end{figure}

\begin{figure}
\plotfiddle{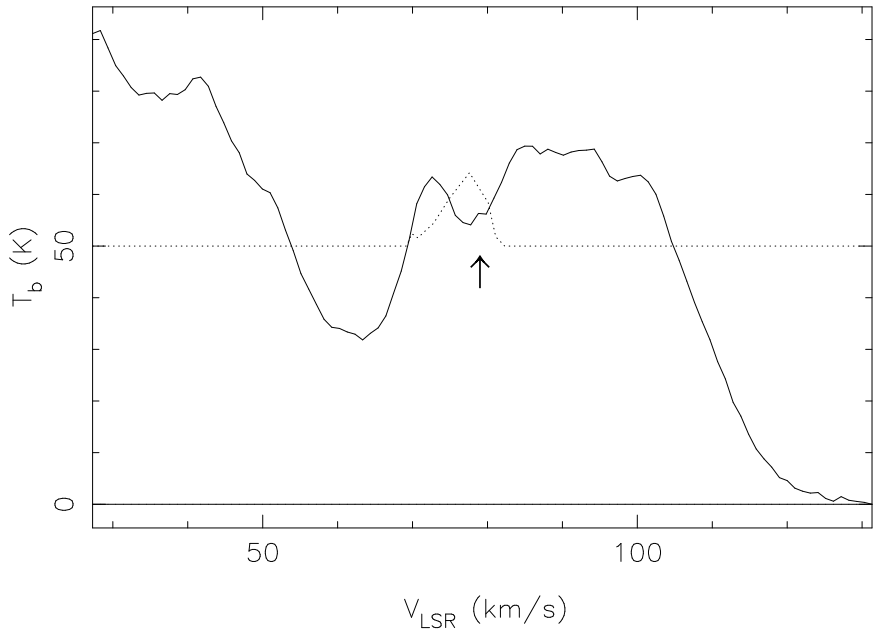}{3in}{0}{50}{50}{-175}{0}
\caption{Spectra of 21cm HI and 1720~MHz OH toward the center of G28.17+0.05.
The 1720~MHz OH emission has been scaled by an arbitrary factor so that its
structure and position can be compared with the 21cm HI emission.  The arrow
marks the center of the HI self-absorption.}
\end{figure}

\section{OH}
The core of G28.17+0.05 is associated with anomalously excited 1720~MHz OH
emission (see Figure 2) (Turner 1982).  
The OH has the same line width as the HI but is blue-shifted by 
$\sim 2.5~{\rm km~s^{-1}}$ relative to the HI.  This OH emission is
brightest toward the core of the cloud.  
 
\section{$^{12}$CO and $^{13}$CO}
Both $^{12}$CO and $^{13}$CO have been observed toward G28.17+0.05.  The $^{12}$CO
emission lines (from Sanders et al. 1986) are seen to have an enhancement 
toward the core.  This indicates that there is at least
$10^5~{\rm M_{\sun}}$ contained in molecular material within G28.17+0.05.
Observations from the BU-FCRAO Inner Ring 
Survey indicate that both $^{12}$CO and $^{13}$CO are self-absorbed toward the
core of G28.17+0.05 (Jackson \& Bania, private communication).  These 
observations give a core temperature of ${\rm T \leq 10~K}$.
Furthermore,
the FCRAO data indicate that the $^{12}$CO and $^{13}$CO line widths are also
the same as the HI line widths. The dominant
mass of G28.17+0.05 is almost certainly molecular.

\section{Star Formation}
There is very little evidence of star formation in G28.17+0.05 
(Myers et al. 1986; Minter et al. 2001) which is unusual for a 
cloud of this size.  There 
is an association of the $^{13}$CO and an MSX Dark Cloud toward G28.17+0.05
(Jackson \& Bania, private communication)
implying regions of extremely high density, perhaps in 
the first stages of star formation.

\section{Conclusions}
There are several indications that G28.17+0.05 is a young cloud:
(1) The emissions lines of HI and OH have a similar width and are
dominated by turbulent motions, suggesting that the turbulence in
G28.17+0.05 has not yet had time to damp.
(2) The lack of star formation.
(3) The large amount of atomic matter in G28.17+0.05.

We suspect that G28.17+0.05 may the first example of a cloud caught in
the transition phase from atomic to molecular as the cloud encounters a
spiral arm shock (Minter et al. 2001).

Since anomalously excited 1720~MHz OH
emission is quite abundant throughout the inner galaxy (Turner 1982), and this
part of the galaxy has not been observed in HI with great detail until
very recently, there could be many more of these atomic/molecular clouds
that have yet to be found or identified.

The VGPS (Taylor et al. 2002) will be extremely valuable for studying this
object, and may reveal many others like it throughout the Galaxy.

\end{document}